\documentclass[%
reprint,
superscriptaddress,
nofootinbib,
amsmath,amssymb, 
aps,
prd,
noeprint,
floatfix,
longbibliography
]{revtex4-2}
\usepackage{color}
\usepackage[utf8]{inputenc}
\usepackage{graphicx}
\usepackage{dcolumn}
\usepackage{multirow}
\usepackage{bm}
\usepackage{upgreek}
\usepackage{siunitx}
\usepackage{booktabs}

\definecolor{greenish}{rgb}{0.09, 0.45, 0.27}

\usepackage{ulem}

\newcommand{\mtwonue}{\mbox{$m^2_\upnu$}}
\newcommand{\mnue}{\mbox{$m_\upnu $}}

\newcommand{\me}{\mbox{$m_\mathrm{e}$}}

\newcommand{\rom}[1]
    {\MakeUppercase{\romannumeral #1}}
\begin{document}

\preprint{APS/123-QED}

\title{Improved eV-scale Sterile-Neutrino Constraints from the Second KATRIN Measurement Campaign}


\newcommand{\berlin}{Institut f\"{u}r Physik, Humboldt-Universit\"{a}t zu Berlin, Newtonstr.~15, 12489 Berlin, Germany}
\newcommand{\bonn}{Helmholtz-Institut f\"{u}r Strahlen- und Kernphysik, Rheinische Friedrich-Wilhelms-Universit\"{a}t Bonn, Nussallee 14-16, 53115 Bonn, Germany}
\newcommand{\cmu}{Department of Physics, Carnegie Mellon University, Pittsburgh, PA 15213, USA}
\newcommand{\cwru}{Department of Physics, Case Western Reserve University, Cleveland, OH 44106, USA}
\newcommand{\etp}{Institute of Experimental Particle Physics~(ETP), Karlsruhe Institute of Technology~(KIT), Wolfgang-Gaede-Str.~1, 76131 Karlsruhe, Germany}
\newcommand{\fulda}{University of Applied Sciences~(HFD)~Fulda, Leipziger Str.~123, 36037 Fulda, Germany}
%
\newcommand{\iap}{Institute for Astroparticle Physics~(IAP), Karlsruhe Institute of Technology~(KIT), Hermann-von-Helmholtz-Platz 1, 76344 Eggenstein-Leopoldshafen, Germany}
\newcommand{\ipe}{Institute for Data Processing and Electronics~(IPE), Karlsruhe Institute of Technology~(KIT), Hermann-von-Helmholtz-Platz 1, 76344 Eggenstein-Leopoldshafen, Germany}
\newcommand{\itep}{Institute for Technical Physics~(ITEP), Karlsruhe Institute of Technology~(KIT), Hermann-von-Helmholtz-Platz 1, 76344 Eggenstein-Leopoldshafen, Germany}
\newcommand{\ppq}{Project, Process, and Quality Management~(PPQ), Karlsruhe Institute of Technology~(KIT), Hermann-von-Helmholtz-Platz 1, 76344 Eggenstein-Leopoldshafen, Germany    }
%
%
\newcommand{\inr}{Institute for Nuclear Research of Russian Academy of Sciences, 60th October Anniversary Prospect 7a, 117312 Moscow, Russia}
\newcommand{\lbnl}{Institute for Nuclear and Particle Astrophysics and Nuclear Science Division, Lawrence Berkeley National Laboratory, Berkeley, CA 94720, USA}
\newcommand{\madrid}{Departamento de Qu\'{i}mica F\'{i}sica Aplicada, Universidad Autonoma de Madrid, Campus de Cantoblanco, 28049 Madrid, Spain}
\newcommand{\mainz}{Institut f\"{u}r Physik, Johannes-Gutenberg-Universit\"{a}t Mainz, 55099 Mainz, Germany}
\newcommand{\mpp}{Max-Planck-Institut f\"{u}r Physik, F\"{o}hringer Ring 6, 80805 M\"{u}nchen, Germany}
\newcommand{\massit}{Laboratory for Nuclear Science, Massachusetts Institute of Technology, 77 Massachusetts Ave, Cambridge, MA 02139, USA}
\newcommand{\mpik}{Max-Planck-Institut f\"{u}r Kernphysik, Saupfercheckweg 1, 69117 Heidelberg, Germany}
\newcommand{\muenster}{Institute for Nuclear Physics, University of M\"{u}nster, Wilhelm-Klemm-Str.~9, 48149 M\"{u}nster, Germany}
\newcommand{\npi}{Nuclear Physics Institute,  Czech Academy of Sciences, 25068 \v{R}e\v{z}, Czech Republic}
\newcommand{\unc}{Department of Physics and Astronomy, University of North Carolina, Chapel Hill, NC 27599, USA}
\newcommand{\washington}{Center for Experimental Nuclear Physics and Astrophysics, and Dept.~of Physics, University of Washington, Seattle, WA 98195, USA}
\newcommand{\wuppertal}{Department of Physics, Faculty of Mathematics and Natural Sciences, University of Wuppertal, Gau{\ss}str.~20, 42119 Wuppertal, Germany}
\newcommand{\saclay}{IRFU (DPhP \& APC), CEA, Universit\'{e} Paris-Saclay, 91191 Gif-sur-Yvette, France }
\newcommand{\tum}{Technische Universit\"{a}t M\"{u}nchen, James-Franck-Str.~1, 85748 Garching, Germany}
\newcommand{\uhd}{Institute for Theoretical Astrophysics, University of Heidelberg, Albert-Ueberle-Str.~2, 69120 Heidelberg, Germany}
\newcommand{\tunl}{Triangle Universities Nuclear Laboratory, Durham, NC 27708, USA}
%
%
\newcommand{\ornl}{Also affiliated with Oak Ridge National Laboratory, Oak Ridge, TN 37831, USA}
%
%
%

\affiliation{\iap}
\affiliation{\ipe}
\affiliation{\inr}
\affiliation{\muenster}
\affiliation{\etp}
\affiliation{\itep}
\affiliation{\tum}
\affiliation{\mpp}
\affiliation{\unc}
\affiliation{\tunl}
\affiliation{\lbnl}
\affiliation{\wuppertal}
\affiliation{\madrid}
\affiliation{\washington}
\affiliation{\npi}
\affiliation{\massit}
\affiliation{\cmu}
\affiliation{\saclay}
\affiliation{\mpik}
\affiliation{\berlin}
\affiliation{\uhd}
\affiliation{\mainz}


\author{M.~Aker}\affiliation{\iap}
\author{D.~Batzler}\affiliation{\iap}
\author{A.~Beglarian}\affiliation{\ipe}
\author{J.~Behrens}\affiliation{\iap}
\author{A.~Berlev}\affiliation{\inr}
\author{U.~Besserer}\affiliation{\iap}
\author{B.~Bieringer}\affiliation{\muenster}
\author{F.~Block}\affiliation{\etp}
\author{S.~Bobien}\affiliation{\itep}
\author{B.~Bornschein}\affiliation{\iap}
% ...

\collaboration{KATRIN Collaboration}\noaffiliation

\date{\today}

\begin{abstract} 
We present the results of the light sterile neutrino search from the second KATRIN measurement campaign in 2019. Approaching nominal activity, \SI{3.76E6}{} tritium $\upbeta$-electrons are analyzed in an energy window extending down to \SI{40}{\electronvolt} below the tritium endpoint at $E_0=\SI{18.57}{\kilo\electronvolt}$. We consider the $3\nu+1$ framework with three active and one sterile neutrino flavor. The analysis is sensitive to a fourth mass eigenstate $m_4^2\lesssim\SI{1600}{\electronvolt\squared}$ and active-to-sterile mixing $|U_{e4}|^2 \gtrsim \SI{6e-3}{}$. As no sterile-neutrino signal was observed, we provide improved exclusion contours on $m_4^2$ and $|U_{e4}|^2$ at \SI{95}{\percent} C.L. Our results supersede the limits from the Mainz and Troitsk experiments. Furthermore, we are able to exclude the large $\Delta m_{41}^2$ solutions of the reactor antineutrino and gallium anomalies to a great extent. The latter has recently been reaffirmed by the BEST collaboration and could be explained by a sterile neutrino with large mixing. While the remaining solutions at small $\Delta m_{41}^2$ are mostly excluded by short-baseline reactor experiments, KATRIN is the only ongoing laboratory experiment to be sensitive to relevant solutions at large $\Delta m_{41}^2$ through a robust spectral shape analysis.
\end{abstract}

\keywords{Suggested keywords}
\maketitle
\section{\em Introduction}
The Karlsruhe Tritium Neutrino (KATRIN) experiment~\cite{KDR2004,KATRIN:2021dfa} is designed to determine the absolute neutrino-mass scale via the kinematics of single $\upbeta$-decay of molecular tritium 
\begin{equation}
\mathrm{T}_2 \rightarrow {}^3\mathrm{HeT}^+ + e^- + \bar{\nu}_e
\end{equation}
with an unprecedented sensitivity \SI{0.2}{\electronvolt} (\SI{90}{\percent} C.L.) after five years of measurement time~\cite{KDR2004}. This is achieved by measuring the integrated $\upbeta$-electron spectrum in a narrow energy interval around the tritium endpoint at $E_0=\SI{18.57}{k\electronvolt}$. The three known neutrino mass eigenstates $m_i$ lead to a reduction of the maximal observed electron energy as well as to a slight spectral shape distortion. As the mass-squared splittings $\Delta m_{ij}^2$ are known to be small compared to the energy resolution of KATRIN~\cite{ParticleDataGroup:2020ssz}, the observable is the squared effective electron antineutrino mass\footnote{In this work we use the convention $c=1$.}
\begin{equation}
m_\nu^2 = \sum_{i=1}^{3} |U_{ei}|^2 m_i^2 ,
\end{equation}
where $U_{ei}$ are the elements of the Pontecorvo-Maki-Nagawa-Sakata (PMNS) matrix~\cite{ParticleDataGroup:2020ssz}. To date, KATRIN provides the most stringent upper limit from a direct measurement of $m_\nu\leq \SI{0.8}{\electronvolt}$ (\SI{90}{\percent} C.L.)~\cite{Aker:2021gma} following its limit of $m_\nu\leq \SI{1.1}{\electronvolt}$ (\SI{90}{\percent} C.L.)~\cite{KATRIN:2019yun} from the first measurement campaign.

Using the same data sets, this work investigates the existence of a fourth neutrino mass eigenstate $m_4$ with active-to-sterile mixing\footnote{Here, we use active-to-sterile mixing as shorthand for the mixing of the electron flavor eigenstate with a fourth mass eigenstate.} $|U_{e4}|^2$. Based on the measured width of the $Z^0$ resonance, it is well established that there are only three light active neutrinos~\cite{ALEPH:2005ab}. Therefore, $m_4$ would be mostly composed of a sterile neutrino flavor that does not participate in the weak interaction. The signature of a sterile neutrino in KATRIN is a kink-like spectral distortion, that is most prominent at electron energies around $E_0 - m_4$. Since the neutrino-mass campaigns focus on measuring the $\upbeta$-spectrum in the vicinity of the endpoint, our sterile-neutrino analysis is restricted to light sterile neutrinos at the eV scale. An extension of the measured energy range to the complete tritium $\upbeta$-decay spectrum would offer the opportunity to search for keV-scale sterile neutrinos with KATRIN. The TRISTAN project plans to extend the KATRIN setup with a novel detector system after completion of the neutrino-mass campaigns to handle the high rates involved in such a measurement~\cite{Mertens_2019}.

Light sterile neutrinos are motivated by accumulating anomalies in short-baseline neutrino oscillation experiments. Studying the appearance of $\nu_e$ from an accelerator $\nu_\mu$ beam, the results from LSND and MiniBooNE suggest evidence for non-standard neutrino oscillations involving a new neutrino mass eigenstate~\cite{PhysRevD.64.112007, Aguilar_Arevalo_2018}. 
Moreover, the gallium anomaly (GA), observed by both GALLEX and SAGE and recently reaffirmed by BEST, describes a $\nu_e$ deficit from $^{37}$Ar and $^{51}$Cr electron capture decays~\cite{Hampel:1997fc,Abdurashitov:2009tn,barinov2021results}. Additionally, a significant discrepancy between the predicted and observed $\bar{\nu}_e$ flux from nuclear reactors, denoted as the reactor antineutrino anomaly (RAA), has been found~\cite{Mention:2011rk}. These neutrino disappearance anomalies could be explained with the existence of a $\gtrsim \SI{1}{\electronvolt}$ sterile neutrino. However, due to difficulties in assessing systematic uncertainties, these anomalies are debated. The recent $>3\sigma$ claim on a $\sim\SI{2.7}{\electronvolt}$ sterile neutrino with large active-to-sterile mixing by the Neutrino-4 $\bar{\nu}_e$ reactor experiment kindled a controversial discussion~\cite{NEUTRINO-4:2018huq}. 

KATRIN offers an approach complementary to short-baseline neutrino oscillation experiments, to search for light sterile neutrinos. Being sensitive to the same parameters that could explain the GA and the RAA, KATRIN is able to probe both anomalies in an independent way. This was first demonstrated with data from the first science run (KNM1), in which parts of the parameter space covered by the sterile-neutrino anomalies could be constrained~\cite{KATRIN:2020dpx}. Here, we present the results of the second measurement campaign (KNM2) as well as the combination of the first two campaigns, improving with respect to the previous KATRIN exclusion bounds.
\section{\em Experimental setup}
\begin{figure*}[htp]
\centering
\includegraphics[width = .9\textwidth]{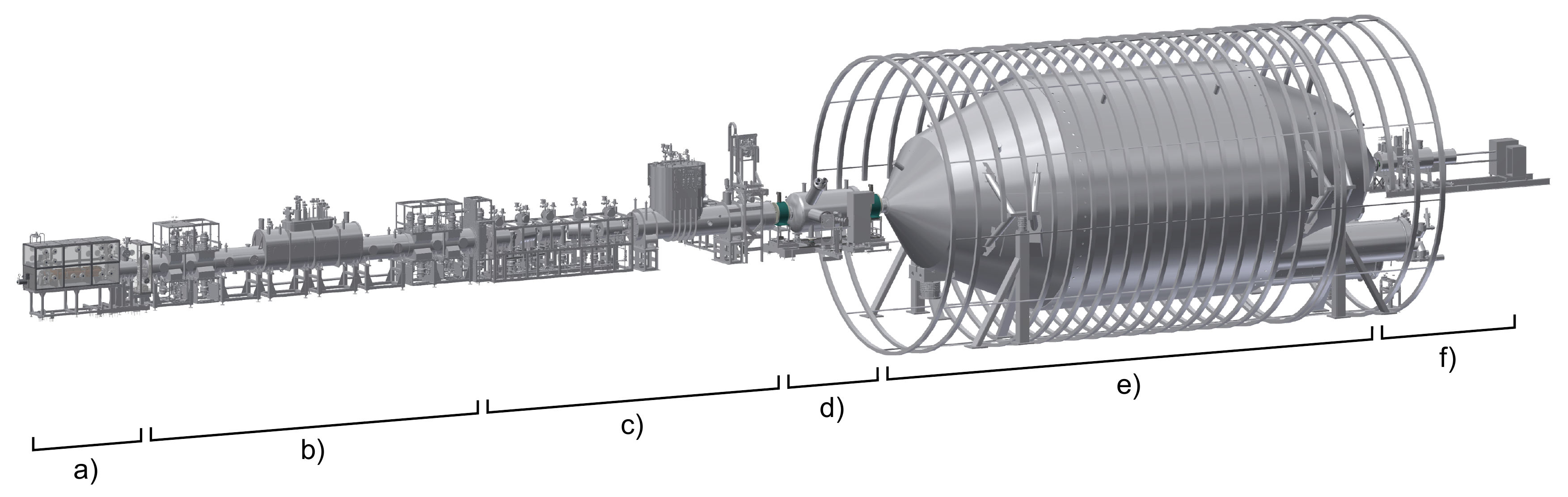}
\caption{The KATRIN beamline is composed of six main components: a) rear wall and electron gun, b) windowless gaseous tritium source, c) transport and pumping section, d) pre-spectrometer, e) main spectrometer, and f) focal plane detector.}
\label{fig:KatrinBEamline}
\end{figure*}

The prerequisites to measure the subtle imprints of $m_\nu^2$ and $m_4^2$ in the tritium $\upbeta$-decay spectrum are a high source activity $\mathcal{O}(\SI{E11}{Bq})$, a low background rate $\mathcal{O}(\SI{0.1}{cps})$, and an eV-scale energy resolution. 
To achieve this, the \SI{70}{m} long KATRIN experiment combines a Windowless Gaseous Tritium Source (WGTS) with a high-precision Magnetic Adiabatic Collimation and Electrostatic (MAC-E) filter~\cite{KDR2004, KATRIN:2021dfa}.
An overview of the experimental setup is displayed in Fig.~\ref{fig:KatrinBEamline}.
High-purity molecular tritium gas ($>\SI{95}{\percent}$) is continuously injected at the center of the \SI{10}{m} long WGTS, keeping the source activity stable at the \SI{0.2}{\percent} level. To minimize thermal gas motion, the WGTS is cooled to \SI{30}{K}~\cite{Grohmann:2013ifa}, which is above the freezing point of T$_2$. Moving from the source towards the spectrometer, the partial tritium pressure is reduced by more than 14 orders of magnitude in the transport and pumping section. This reduction is achieved through a differential and a cryogenic pumping section \cite{MARSTELLER2021109979}. Electrons from the source are magnetically guided through the setup~\cite{Arenz:2018jpa}, starting in a magnetic field of $B_\mathrm{source}=\SI{2.5}{T}$.

Applying the MAC-E-filter principle~\cite{Lobashev:1985mu, Picard1992}, the energies of the electrons are evaluated in the spectrometer section consisting of the pre- and main spectrometers. By gradually reducing the magnetic field towards the analyzing plane to $B_\mathrm{ana}=\SI{6.3e-4}{T}$ in the main spectrometer, the momenta of all electrons are adiabatically collimated. At the exit of the main spectrometer, the flux tube is tapered by the magnetic field $B_\mathrm{max}=\SI{4.2}{T}$ of the superconducting pinch magnet. Both spectrometers act as electrostatic high-pass filters, allowing only electrons with kinetic energies larger than the applied retarding energy $qU$ to be transmitted. Given the magnetic field configuration, the filter width at the endpoint is $\Delta E =\SI{18.6}{keV} \cdot (B_\mathrm{ana}/B_\mathrm{max})=\SI{2.78}{eV}$ with a maximum accepted angle of $\theta_\mathrm{max} = \sqrt{B_\mathrm{source}/B_\mathrm{max}} = \SI{50.4}{\degree}$.
Electrons transmitted through the main spectrometer are finally counted as a function of $qU$ at the focal plane detector (FPD)~\cite{Amsbaugh:2014uca}. The FPD is a monolithic silicon PIN-diode segmented into 148 pixels of equal area. At the upstream end of the experiment, non-transmitted electrons are eventually absorbed by a gold-plated rear wall. To homogenize the source electric potential, the rear wall is biased to a voltage of $\mathcal{O}(\SI{100}{\milli\volt})$, compensating for the intrinsic potential difference between the rear wall and the beam-tube surface.
\section{\em The KNM2 measurement campaign}
In the following, we describe the second measurement campaign (KNM2): Section~\ref{sec:measurement1} addresses the operating conditions of the KATRIN experiment as well as the measurement procedure. The data selection is presented in Sec.~\ref{sec:dataselection} and the data combination in Sec.~\ref{sec:datacombi}.
\subsection{\em Measurement of the integral tritium spectrum}\label{sec:measurement1}
The second high-purity tritium campaign was conducted in October and November of 2019. We achieved a source activity of \SI{9.5e10}{\becquerel}, improving with respect to our first measurement campaign by a factor of $3.8$. The operation with $\rho d = \SI{4.2e17}{molecules\per\centi\meter\squared}$, close to the designed nominal column density, was possible as we are no longer limited to operation in the `burn-in' configuration~\cite{Aker_2021}. 
By continuously monitoring the gas composition with a laser Raman system, we determined the isotopic tritium purity as $\epsilon_T=\SI{98.7}{\percent}$~\cite{LARA:20174827}. The background rate was reduced by \SI{25}{\percent} to \SI{220}{mcps} following an improvement of the vacuum conditions in the main spectrometer~\cite{Aker:2021gma}.
We recorded the integrated $\upbeta$-spectrum by repeatedly measuring the count rate $R_\mathrm{data}(qU_i)$ at 39 different retarding energies in a range of $[E_0 - \SI{300}{\electronvolt}$, $E_0 + \SI{135}{\electronvolt}]$. One measurement at a given $qU_i$ is called a scan-step and lasts between \SI{17}{\second} and \SI{576}{\second}. The ensemble of all 39 scan steps, referred to as a \textit{scan}, had a duration of \SI{2}{\hour}. The measurement time distribution within a scan is displayed in Fig.~\ref{fig:spectrum} c) for the analyzed energy range. It was optimized with respect to the neutrino mass sensitivity prior to the measurement campaign. The scan-steps above $E_0$ are used to determine the constant background rate.
\subsection{\em Data selection}\label{sec:dataselection}
In this work we limit the analysis to the energy range $[E_0 - \SI{40}{\electronvolt}$, $E_0 + \SI{135}{\electronvolt}]$, shown in Fig.~\ref{fig:spectrum}, in which the measurement is dominated by statistical uncertainties. Applying data-quality criteria identical to~\cite{Aker:2021gma} result in 361 out of 397 recorded scans selected for the analysis. The rejection of the other scans is based on insufficient gas-composition data as well as corrupted settings of the high voltage. The accepted measurement data amount to a total scan time of \SI{744}{\hour}.
From the available 148~pixels of the FPD, we select 117 pixels for the analysis. The rejected pixels do not satisfy quality requirements, as they exhibit an increased noise level, a broadened energy resolution or suffer from misalignment of the beam-line with respect to the magnetic flux tube. 
\subsection{\em Data combination}\label{sec:datacombi}
The sub-ppm precision level of the high-voltage (HV) system, combined with a source-potential stability of $\sigma<\SI{80}{\milli\volt}$, allows us to combine the selected scans into one effective spectrum with averaged HV values $\langle qU\rangle_i$. The temporal source-potential variations are incorporated in our model as an energy broadening in the final-state distribution (see Sec.~\ref{sec:model}).
In addition, the excellent homogeneity of the electric ($\sigma<\SI{41}{\milli\volt}$) and magnetic ($\sigma<\SI{E-6}{\tesla}$) fields in the analyzing plane, as well as consistent pixel characteristics, justify the combination of all selected pixels to one effective detector area (uniform fit). Possible radially-dependent source-potential variations, estimated with a fit that allows for a radial-dependent effective endpoint, are found to be negligible ($<\SI{100}{\milli\electronvolt}$)~\cite{Aker:2021gma}.
The combined data spectrum that is used for the final spectral fit is displayed in Fig.~\ref{fig:spectrum} a).
In the \SI{40}{\electronvolt} energy range below the endpoint, \SI{3.76E6}{} tritium $\upbeta$-electrons and \SI{0.41E6}{} background electrons were counted. 
While the number of background electrons has only a slight retarding energy dependence, the number of signal electrons increases steeply with decreasing $qU$. Therefore, we observe $qU$-dependent signal-to-background ratios of $(235, 85, 20, 1)$ at $(\num{40}, \num{30}, \num{20}, \num{10})$ \si{\electronvolt} below the endpoint.
\section{Experimental modeling}\label{sec:model}
The following section will give an overview of the analytical model describing the measured $\upbeta$-spectrum. More details can be found in Ref.~\cite{Kleesiek:2018mel}.
\subsection{\em Differential spectrum}
The differential spectrum $R_\upbeta(E)$ of the super-allowed tritium $\upbeta$-decay can be derived using Fermi's golden rule:
\begin{equation}\label{eq:diff-spectrum}
  \begin{split}
    R_\upbeta(E, \mtwonue) =& \frac{G_\mathrm{F}^2 \cdot \cos^2\Theta_C}{2\pi^3}\left|M_\mathrm{nucl}\right|^2 \cdot F\left(Z', E\right) \\
    & \cdot \left(E+\me\right)\cdot\sqrt{\left(E+\me\right)^2 - m_\mathrm{e}^2}\\
    & \cdot \sum_{f} \zeta_f\varepsilon_f(E)\sqrt{(\varepsilon_f(E))^2 -\mtwonue} \\
    & \cdot \Theta(\varepsilon_f(E)-\mnue).
  \end{split}
\end{equation}
Here, $G_\mathrm{F}$ is the Fermi constant, $\Theta_C$ is the Cabibbo angle, and $\left|M_\mathrm{nucl}\right|^2$ is the energy-independent nuclear matrix element. $F\left(Z', E\right)$ is the Fermi function with $Z'=2$ for the helium daughter nucleus.
The neutrino energy $\varepsilon_f$ is given by $\varepsilon_f(E) = E_0 - E - V_f$, with the kinematic endpoint $E_0$.
$E$ represents the kinetic energy of the electron, and $\me$ is the electron mass.
Rotational, vibrational and electronic excitations of parent and daughter molecules are taken into account by summing over all molecular final state energies $V_f$ with probabilities $\zeta_f$~\cite{Saenz2000, Doss2006, Aker_2021}.

Further corrections are taken into account in the spectrum such as the Doppler broadening due to thermal motion of molecules in the WGTS and theoretical corrections~\cite{Kleesiek:2018mel}. Spatial and temporal variations of the source potential are incorporated as an effective energy broadening ($\sigma^2=\SI[separate-uncertainty=true]{12.4 \pm 16.1e-3}{\electronvolt\squared}$) in the final-state distribution and as a shift in the energy loss function ($\Delta_\text{P}=\SI[separate-uncertainty=true]{0 \pm 61}{meV}$)~\cite{Aker_2021}.

\begin{figure}[htp]
    \centering
    \includegraphics[width=\columnwidth]{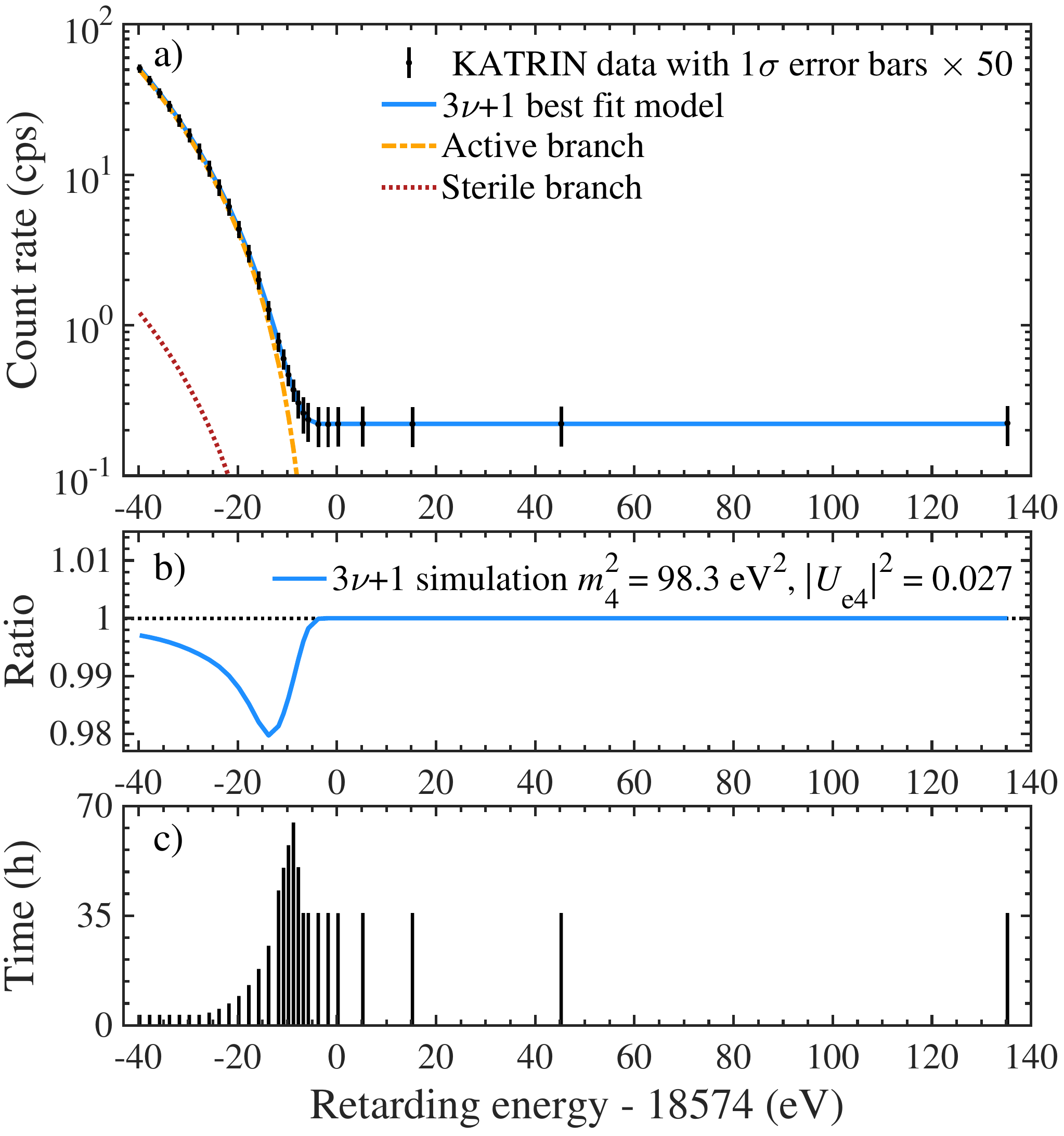}
    \caption{a) KATRIN data and $3\upnu+1$ best-fit model of the second measurement campaign with statistical uncertainties. The error bars are scaled by a factor of 50 to improve visibility. The best-fit model in the $3\upnu+1$ framework comprises signal contributions from both the active ($m_\nu^2 = \SI{1.1}{\electronvolt\squared}$) and sterile ($m_4^2 = \SI{98.3}{\electronvolt\squared}$, $|U_{e4}|^2 = 0.027$) branches. This result improves only insignificantly ($\Delta\chi^2 = 2.5$) upon the framework without a sterile neutrino. The best-fit uncertainties on the two sterile parameters are incorporated in the exclusion contour in Sec.~\ref{sec:ResultFreemNuSq}. The $m_\nu^2$ uncertainty is discussed in Sec.~\ref{sec:result_numass}. b) Illustration of the sterile-neutrino signal in KATRIN. The ratio of two simulated spectra without statistical fluctuations in the $3\upnu$+1 and $3\upnu$ frameworks is shown. The sterile-neutrino signature is a kink-like structure, most prominent at retarding energies around $qU \approx E_0 - m_4$. c) The integrated spectrum is measured by successively applying different high voltages to the main spectrometer. The cumulative time spent at each retarding energy forms the measurement time distribution.}
    \label{fig:spectrum}
\end{figure}

In our sterile-neutrino analysis, we extend the standard $\upbeta$-decay model by a sterile decay branch associated with a fourth neutrino mass eigenstate $m_4^2$ and active-to-sterile mixing $|U_{e4}|^2$. The decay spectrum in Eq.~\eqref{eq:diff-spectrum} is replaced by
\begin{equation}\label{eq:3nup1}
  \begin{split}
    R_\upbeta(E,m_\upnu^2,m_4^2,|U_{e4}|^2) =& \\(1-|U_{e4}|^2) \cdot R_\upbeta(E,m_\upnu^2) +  |U_{e4}|^2 \cdot R_\upbeta(E,m_4^2)
  \end{split}
\end{equation}
with the extended $4\times4$ unitary PMNS mixing matrix $U$. The simulated imprint of an eV-scale fourth mass eigenstate with nonzero mixing is illustrated in Fig.~\ref{fig:spectrum} b).
\subsection{\em Experimental Response function}
The response function $f(E,qU_i)$ describes the probability of transmission through the beamline ($\mathcal{T}(E,\theta,U)$) and energy losses $\epsilon$ due to inelastic scattering in the source. In this work the same description of the response function as in Ref.~\cite{Aker:2021gma} is used. 
The response function is given by
\begin{equation}\label{eq:response}
  \begin{split}
    f(E,qU_i) =& \int_{\epsilon=0}^{E-qU_i}\int_{\theta=0}^{\theta_\mathrm{max}}\mathcal{T}(E-\epsilon,\theta,U_i) \\
    &\cdot \sin\theta \cdot \sum_{s} P_s(\theta)f_s(\epsilon)\mathrm{d}\theta\mathrm{d}\epsilon.
  \end{split}
\end{equation}
The integrated transmission probability $T(E,\theta,U)$ is written as
\begin{widetext}
  \begin{equation}\label{eq:transmission}
  \begin{split}
    T(E,\theta,U) =& \int_{\theta=0}^{\theta_\mathrm{max}}\mathcal{T}(E,\theta,U_i) \sin\theta\mathrm{d}\theta \\ =&
    \begin{cases}
    0  & , E - qU < 0 \\
    1 - \sqrt{1-\frac{E-qU}{E}\frac{B_\mathrm{source}}{B_\mathrm{ana}}\frac{2}{\gamma+1}} & , 0 \leq E - qU \geq \Delta E\\
    1 - \sqrt{1-\frac{B_\mathrm{source}}{B_\mathrm{max}}} & , E - qU > \Delta E\\
    \end{cases}
  \end{split}
  \end{equation}
\end{widetext}
with the Lorentz factor $\gamma$. The transmission function determines the filter width of the main spectrometer, which depends on the magnetic fields in the source $B_\mathrm{source}$, the analyzing plane $B_\mathrm{ana}$, and the maximum magnetic field $B_\mathrm{max}$.
The energy loss due to inelastic scattering with tritium in the source is described by the energy loss function $f_s(\epsilon)$ for $s$-fold scattering and the associated scattering probability $P_s(\theta)$. The latter depends on the emission angle $\theta$, which is defined as the initial polar angle of the electron momentum, relative to the magnetic field. More details can be found in~\cite{Kleesiek:2018mel}. 
\subsection{\em Spectrum prediction}
The integrated spectrum rate $R_\mathrm{model}(qU_i)$ that is measured at the detector is predicted by
\begin{equation}\label{eq:int-spectrum}
\begin{split}
  R_\mathrm{model} (qU_i) =& A_\mathrm{S} N_\mathrm{T}\int_{qU_i}^{E_0} R_\upbeta(E)f(E,qU_i) \mathrm{d}E\\
                & + R_\mathrm{bg}(qU_i).
\end{split}
\end{equation}
Here the differential $\upbeta$-spectrum of the tritium decay $R_\upbeta(E)$ from Eq.~\eqref{eq:3nup1} is convolved with the response function $f(E,qU_i)$ from Eq.~\eqref{eq:response}. The integrated spectrum is multiplied by two normalization factors, ($N_\mathrm{T}$ and $A_\mathrm{S}$). Firstly, $N_\mathrm{T}$ scales the spectrum to the number of tritium molecules within the flux tube quantified by experimental determination, multiplied with the accepted solid angle and the detection efficiency. Secondly, the amplitude of the tritium signal, $A_\mathrm{S}$, is considered a free fit parameter as the prediction of the absolute decay rate is not accurate enough.
The background rate $R_\mathrm{bg}(qU_i)$ comprises three components: A dominant flat background base $R_\mathrm{bg}^\mathrm{base}$, a hypothetical retarding-potential-dependent background $R_\mathrm{bg}^{qU}(qU_i)$, and a small background contribution from electrons stored in the Penning trap between the pre- and main spectrometers $R_\mathrm{bg}^\mathrm{Penning}(t_i)$~\cite{Aker:2021gma}.
\begin{equation}
  \begin{split}
    R_\mathrm{bg}(qU_i) =&  R_\mathrm{bg}^\mathrm{base} + R_\mathrm{bg}^{qU}(qU_i) \\ &+ R_\mathrm{bg}^\mathrm{Penning}(t_i(qU_i))
  \end{split}
\end{equation}
The Penning background increases with the mean scan-step duration $t_i(qU_i)$ at a given retarding energy setting $qU_i$.
The retarding potential dependence of the background is constrained to $\SI[separate-uncertainty=true]{0.0 \pm 4.7}{mcps/keV}$. The scan-step-duration dependence of the Penning background is constrained to $\SI[separate-uncertainty=true]{3.0 \pm 3.0}{\micro cps/s}$. These constraints were determined from independent analyses~\cite{Aker:2021gma}.
\section{\em Data Analysis}
In the following, we describe the analysis methods of our sterile-neutrino search. Section~\ref{sec:sterileAnalysis} is dedicated to the inference of the sterile-neutrino parameters ($m_4^2$ and $|U_{e4}|^2$) from the tritium $\upbeta$-spectrum using the grid search technique. In this context the applicability of Wilks' theorem and the blinding procedure to ensure a robust and unbiased analysis are discussed. Moreover, the propagation of systematic uncertainties using the covariance matrix approach is presented. The procedure of generating the covariance matrices, as well as their application, is described in Sec.~\ref{sec:systematics}. Furthermore, the relative contribution of different systematics to the total uncertainty budget is given.
\subsection{\em Sterile-neutrino analysis} \label{sec:sterileAnalysis}
Our sterile-neutrino analysis aims to infer two physics parameters of interest: $m_4^2$ and $|U_{e4}|^2$. Additionally, the four original fit parameters ($m_\upnu^2$, $E_0$, $A_\mathrm{S}$, $R_\mathrm{bg}^\mathrm{base}$) of the neutrino-mass analysis~\cite{Aker:2021gma} are included as nuisance parameters.
Therefore, the sterile-neutrino constraints are retrieved solely from the \emph{shape} information within the experimental spectrum. To infer the sterile-neutrino parameters, we perform fits with fixed $(m_4^2, |U_{e4}|^2)$ pairs, minimizing the standard function
\begin{equation}\label{eq:chi2}
  \begin{split}
\chi^2(\vec{\xi}) =\\ (\vec{R}_\textrm{data}-\vec{R}_\textrm{model}(\vec{\xi})) \ C^{-1} \ (\vec{R}_\textrm{data}-\vec{R}_\textrm{model}(\vec{\xi}))^\top
 \end{split}
\end{equation}
with respect to all other fit parameters for each pair. Here $\vec{\xi}$ denotes the parameter set $(\mtwonue, m_4^2, |U_{e4}|^2, E_0, A_\mathrm{S}, R_\mathrm{bg}^\mathrm{base})$. The elements of the vectors $\vec{R}$ give the rates at different retarding potentials for data and model, respectively. Statistical as well as systematic uncertainties are incorporated in the covariance matrix $C$, which is described in detail in Sec.~\ref{sec:systematics}.

A $(50 \times 50)$ logarithmically-spaced grid over the parameters $m_4^2$ and $|U_{e4}|^2$ $(m_4^2 \in [\num{0.1}, \num{1600}] \, \si{\electronvolt\squared}, |U_{e4}|^2 \in [10^{-3}, 0.5])$ was used. $|U_{e4}|^2 = 0.5$ is considered as maximal mixing: If both $m_4^2$ and $\mtwonue$ are unconstrained fit parameters, the active and sterile branch in the decay spectrum (Eq.~\eqref{eq:3nup1}) are interchangeable: $(m_4^2, |U_{e4}|^2) \leftrightarrow (\mtwonue, 1-|U_{e4}|^2)$. Consequently, no additional information can be gained by extending the active-to-sterile mixing beyond 0.5.
Different grid layouts were studied to ensure full convergence of the employed grid. We draw the \SI{95}{\percent} C.L. exclusion contour at
\begin{equation}
\Delta\chi^2 = \chi^2 - \chi^2_\textrm{min} = 5.99
\end{equation}
 following Wilks' theorem~\cite{Wilks:1938dza} for two degrees of freedom. 
 
 We verified the applicability of Wilks' theorem by constructing the $\Delta\chi^2$ probability distribution functions numerically with $1500$ randomized Monte Carlo simulations for the null hypothesis and several different sterile-neutrino hypotheses. The former is, by definition, the boundary physics case. As the null hypothesis can be realized with an infinite number of sterile parameter pairs with $|U_{e4}| = 0$ and arbitrary $m_4^2$, a hypothetical deviation from Wilk's theorem is anticipated to be most prominent here. Short-baseline oscillation experiments observe a deviation from Wilk's theorem, because statistical fluctuations of single data points can likely mimic a sterile neutrino signal~\cite{AgostiniNeumair2020}. In contrast to that, the sterile-neutrino signal in KATRIN manifests itself as a \emph{global} spectral distortion. Therefore, Wilk's theorem is expected to apply for KATRIN. Indeed, we did not observe deviation from the predictions of Wilks' theorem for any of the Monte Carlo truths.  
 
To mitigate human-induced biases, the full analysis chain is first applied to a simulated data set without statistical fluctuations. For each experimental scan, we generate a twin spectrum based on the true experimental parameters assuming no sterile neutrino and vanishing neutrino mass. Only after three independent analysis teams, using different analysis codes, obtained consistent sensitivity estimates, the actual data analysis was performed without any subsequent modifications.
\subsection{\em Systematic uncertainties}\label{sec:systematics}
Since KATRIN is a high-precision experiment, the accurate description of all systematic effects is crucial. Various systematic effects are taken into account, arising at different points along the electron's trajectory from the source to the focal plane detector. We consider the same systematic effects as in the neutrino mass analysis~\cite{Aker:2021gma}.

The uncertainty propagation to the integrated spectrum is conducted with the covariance-matrix approach~\cite{Aker_2021}. Each independent systematic effect can be assessed by a separate covariance matrix $C_j$. The latter is estimated by simulating $\mathcal{O}(10^4)$ tritium spectra, varying the relevant set of parameters associated to a particular systematic effect $j$ according to their joint probability distribution function. Since we perform a \emph{shape-only} analysis, absolute rate uncertainties are eliminated from the covariance matrices by normalizing each sample spectrum to the average statistics. The sum of all covariance matrices $C_j$ and a diagonal matrix describing statistical uncertainties $C_\mathrm{stat}$ comprises the total variances and covariances of the data points:
\begin{equation}
C = \sum_j C_j + C_\mathrm{stat}.
\end{equation}
Diagonal elements of the covariance matrix give the bin-to-bin uncorrelated uncertainties,
 whereas off-diagonal elements are $qU$-dependent correlations in the integrated spectrum. The impact on the model spectrum is then taken into account by applying the covariance matrix to the $\chi^2$-function (Eq.~\eqref{eq:chi2}) in the final fit. Since the covariance matrices are pre-calculated, it is an efficient approach to include many systematic effects simultaneously without demanding more computing time in the fit itself.
 
The analysis of systematic uncertainties is performed on the simulated twin data set with fixed $\mtwonue = \SI{0}{\electronvolt\squared}$. For each systematic effect a grid scan considering only the individual systematic uncertainty on top of the statistical uncertainty was carried out.
For all systematic effects, the change of the contour is small compared to the sensitivity evaluated with only statistical uncertainties.

\begin{figure*}[htp]
\centering
\includegraphics[width=2\columnwidth]{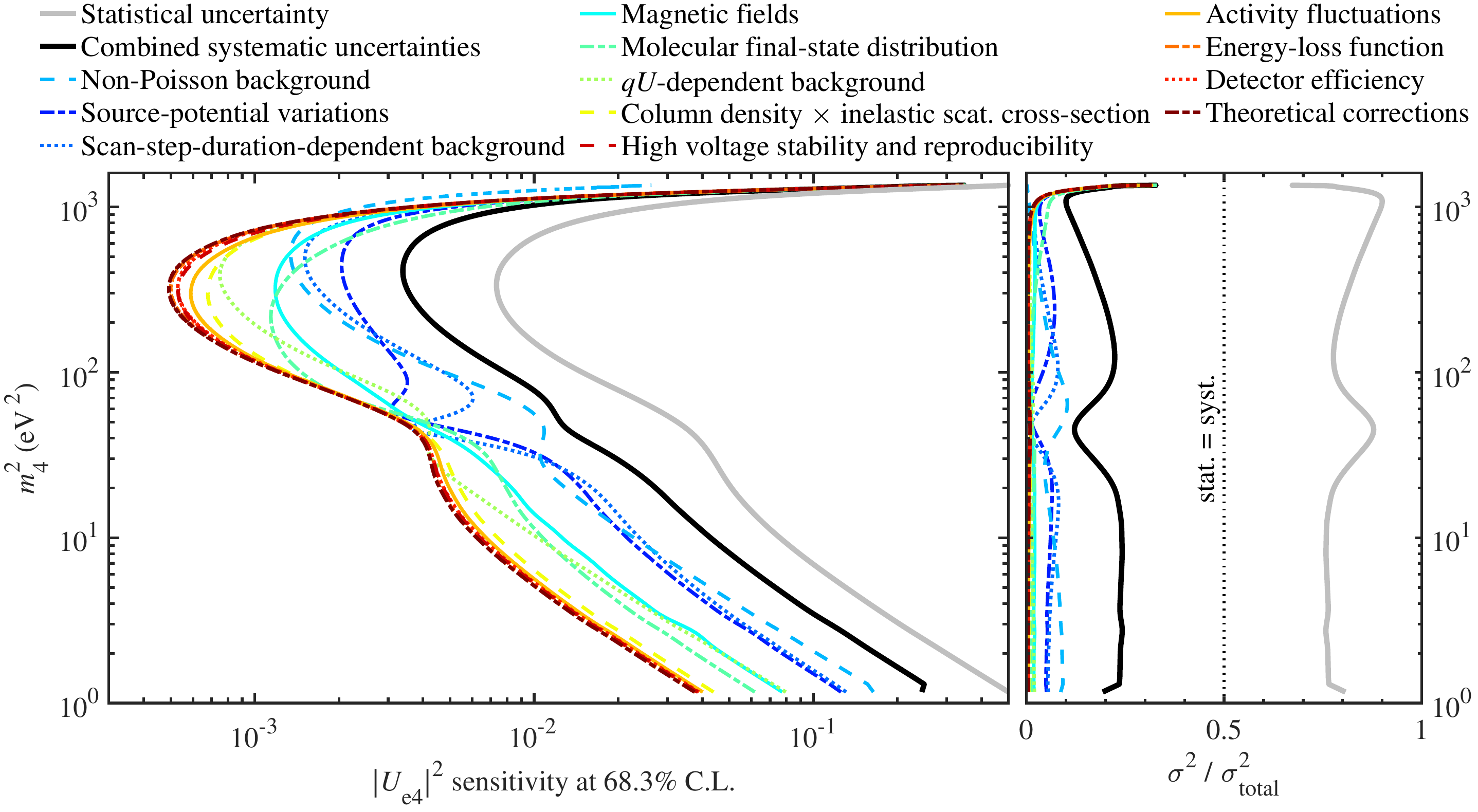}
\caption{Uncertainty breakdown obtained from a simulated twin data set with $\mtwonue = \SI{0}{\electronvolt\squared}$. The left panel shows the systematic-only contours for individual systematic effects and the statistics-only contour. The systematic-only contours are extracted from raster scans described in Sec.~\ref{sec:systematics} using Eq.~\eqref{eq:sysonly}. Only the region $m_4^2>\SI{1}{eV}$ is shown, since there is no sensitivity to the systematic-only contribution in small mass regions. The right panel illustrates the relative contribution of each systematic effect to the total uncertainty budget for all $m_4^2$. The median contributions $(\sigma^2/\sigma_\textrm{total}^2)_\mathrm{median}$ are listed in Tab.~\ref{tab:systematics}. All systematic effects are small compared to the statistical uncertainty, $\sigma_\textrm{syst}^2/\sigma_\textrm{total}^2 \ll 0.5$. The statistical uncertainty even dominates over all systematic uncertainties combined, $\sigma_\textrm{stat}^2/\sigma_\textrm{total}^2 > 0.5$ for all $m_4^2$.}
\label{fig:SysBreakdown_SysOnly}

\end{figure*}

To assess the relative contribution of each systematic effect to the total uncertainty in a more quantitative way, we perform a raster scan for each effect. For fixed values of $m_4^2$, we calculate the 1 $\sigma$ sensitivity on the mixing, $\sigma(|U_{e4}|^2)$, as a function of $m_4^2$. In this case we have only one degree of freedom; therefore the critical $\chi^2$ is reduced to \num{1} at \SI{68.3}{\percent} C.L. This method allows us to assess the \emph{systematic-only} contribution on $|U_{e4}|^2$ using
\begin{equation}\label{eq:sysonly}
\sigma_\textrm{syst} = \sqrt{\sigma^2_\textrm{total}-\sigma^2_\textrm{stat}}.
\end{equation}
This quantity is displayed in the left panel of Fig.~\ref{fig:SysBreakdown_SysOnly} for each systematic effect as a function of $m_4^2$. Moreover, the relative contribution of each systematic variance to the total variance $\sigma_\mathrm{syst}^2/\sigma_\mathrm{total}^2$ is shown in the right-hand panel. In Tab.~\ref{tab:systematics} the median contribution of each systematic effect is summarized, sorted by magnitude. The median is calculated using the same linearly-spaced $m_4^2$ values for all systematic effects.

The analysis of the second KATRIN campaign is statistics-dominated in the analyzed energy range, with $\sigma_\mathrm{stat}^2/\sigma_\mathrm{total}^2 > 0.5$ for all $m_4^2$. For $m_4^2\leq\SI{600}{\electronvolt\squared}$ the systematic effects are dominated by the non-Poisson rate distribution of the background~\cite{fraenkle2020katrin}, the scan-step-duration-dependent background, and source-potential variations. For larger $m_4^2>\SI{600}{\electronvolt\squared}$, all systematic contributions, except for the non-Poisson rate distribution of the background, rapidly increase. The largest systematic contribution for larger $m_4^2$ is given by the molecular final-state distribution. This can be explained by an increased uncertainty on the excited molecular states in this energy region. 
\begin{table*}[!ht]
\small
\begin{center}
\caption{Breakdown of the relative uncertainties on $|U_{e4}|^2$, given as the median $(\sigma^2/\sigma^2_\mathrm{total})_\mathrm{median}$ over all $m_4^2$. The systematic effects are listed in ascending order of the maximal uncertainty $\mathrm{max}(\sigma_\mathrm{syst}^2/\sigma^2_\mathrm{total})$. The systematic uncertainty inputs are those used in the neutrino-mass analysis published in~\cite{Aker:2021gma}. The analysis is dominated by statistical uncertainties ($(\sigma_\mathrm{stat}^2/\sigma^2_\mathrm{total})_\mathrm{median} > 0.5$).}
\label{tab:systematics}
\begin{tabular}{lc}
\toprule
Effect        & $(\sigma^2/\sigma^2_\mathrm{total})_\mathrm{median}$ \\ 
\midrule
\textbf{Statistical}                                & \num{0.86} \\
\midrule
Source-potential variations                         & \num{0.06} \\
Scan-step-duration-dependent background             & \num{0.04} \\
Non-Poisson background                              & \num{0.02} \\
Magnetic fields                                     & \num{0.03} \\
Molecular final-state distribution                  & \num{0.05} \\
$qU$-dependent background                           & \num{0.01} \\
Column density $\times$ inelastic scat. cross section & \num{0.01} \\
Detector efficiency                                 & \num{0.01} \\
Activity fluctuations                               & $<\num{0.01}$ \\
Energy-loss function                                & $<\num{0.01}$ \\
High voltage stability and reproducibility          & $<\num{0.01}$ \\
Theoretical corrections                             & $<\num{0.01}$ \\
\midrule
\textbf{Total systematic uncertainty}                          & \num{0.14} \\
\bottomrule
\end{tabular}
\end{center}
\end{table*}
\section{\em Results}
Here, we report on the results of our light sterile neutrino search. Before presenting the improved exclusion bounds, we address the correlation between active and sterile neutrino branches in the KNM2 tritium $\upbeta$-decay spectrum (Sec.~\ref{sec:NuMassCorrelation}). As the former has a strong influence on the sterile neutrino exclusion bounds, we then distinguish between two analysis cases: In the case \rom{1} analysis (Sec.~\ref{sec:result_fix}) we consider $m_\nu^2 = \SI{0}{\electronvolt\squared}$, whereas the case \rom{2} analysis (Sec.~\ref{sec:ResultFreemNuSq}) employs $m_\nu^2$ as a nuisance parameter. In Sec.~\ref{sec:results_combi} the combined analysis of KNM1 and KNM2 is presented. Lastly, the $m_\nu^2$ sensitivity in the presence of a sterile neutrino is evaluated (Sec.~\ref{sec:result_numass}).
\subsection{Correlation between active and sterile neutrino branches}\label{sec:NuMassCorrelation}
The model spectrum in Eq.~\eqref{eq:3nup1} consists of the weighted sum of two branches: the \emph{active} branch with effective electron antineutrino mass $m_\upnu$ and the \emph{sterile} branch with the fourth mass eigenstate $m_4$. The branches are weighted according to their mixing: $(1-|U_{e4}|^2)$ for the active branch and $|U_{e4}|^2$ for the sterile branch, respectively. Apart from the different neutrino masses and weights, the two branches are mathematically identical. Since $\mtwonue$ is small in the observed data and simulation, the two branches are degenerate in the case of small $m_4^2$ and large mixing $|U_{e4}|^2 \approx 0.5$.

\begin{figure}[htp]
\begin{center}
\includegraphics[width=\columnwidth]{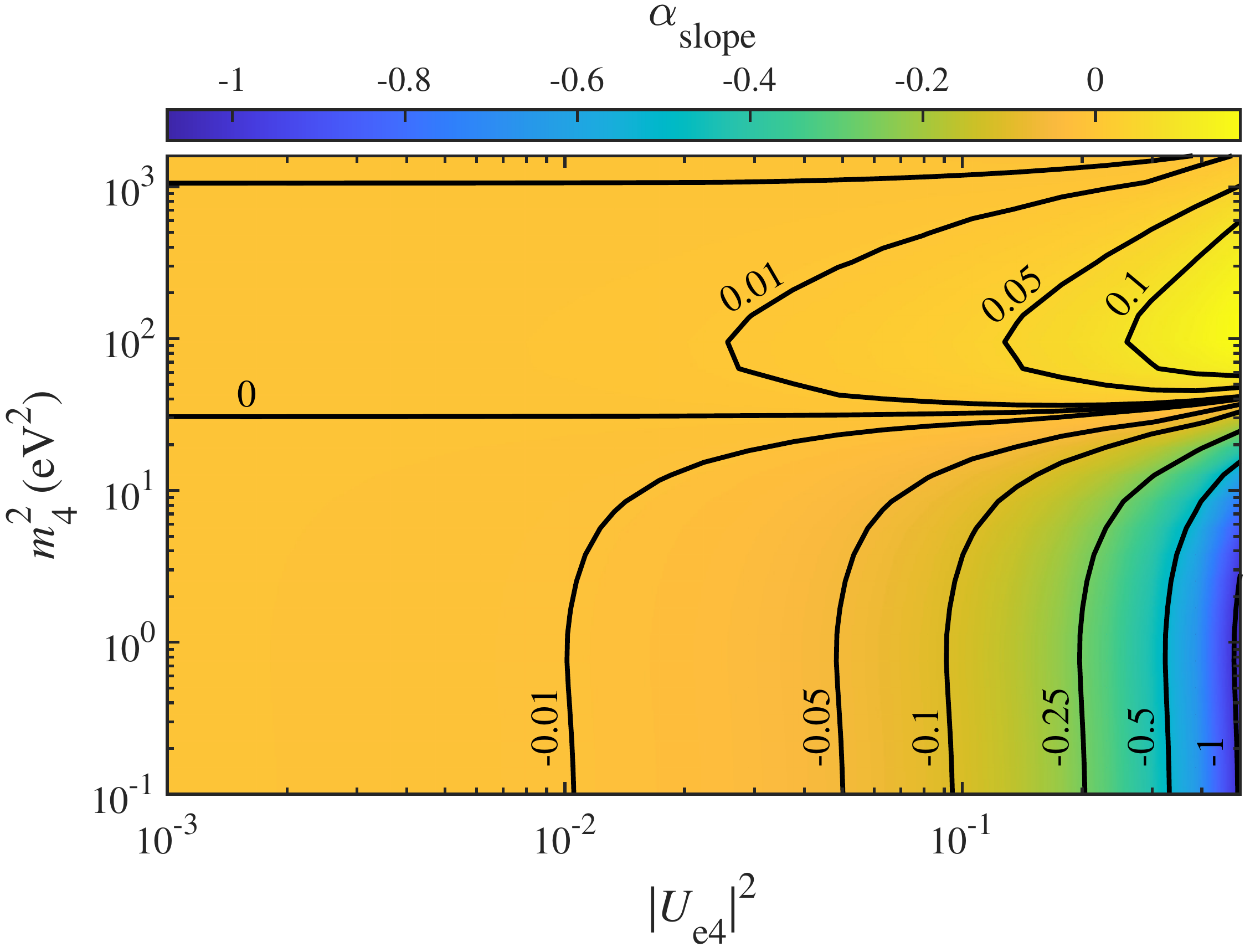}
\caption{\label{fig:NuMassCorrelation} The correlation between active and sterile neutrino mass is approximately a linear slope $m_\nu^2 = \alpha_\textrm{slope} \cdot m_4^2 + \mathrm{const.}$ for various values of $m_4^2$ and $|U_{e4}|^2$ by analyzing simulated spectra. The gradient indicates the magnitude of $\alpha_\textrm{slope}$. For small mixing $|U_{e4}|^2 < 0.01$, we observe small slope values $|\alpha_\textrm{slope}| < 0.01$. For larger mixing we find a strong negative correlation for small $m_4^2 \lesssim \SI{30}{\electronvolt\squared}$ and a weaker positive correlation for larger $m_4^2$.}
\end{center}
\end{figure}

To quantify this relation more generally, we simulate $R_\upbeta$ for several values of $m_4^2, |U_{e4}|^2$ and $\mtwonue=\SI{0}{\electronvolt\squared}$.
Then we perform five fits to each simulated spectrum by varying $m_4^2$ step-wise by $\pm\SI{1}{\electronvolt\squared}$ around the respective MC truths. The fits are optimized with respect to all nuisance parameters, keeping $|U_{e4}|^2$ fixed to its simulated value. For each ($m_4^2,|U_{e4}|^2$)-pair, we determine the approximately linear relationship $m_\nu^2 = \alpha_\textrm{slope} \cdot m_4^2 + \mathrm{const.}$ in the vicinity of the MC truth, which serves as a proxy for the correlation between the two masses. Figure~\ref{fig:NuMassCorrelation} shows $\alpha_\textrm{slope}$ in the $(m_4^2,|U_{e4}|^2$)-parameter space. The smaller the active-to-sterile mixing, the smaller is the contribution of the sterile branch to the simulated spectrum. We find small $|\alpha_\textrm{slope}| < 0.01$ for small $|U_{e4}|^2 < 0.01$. 
For small $m_4^2\lesssim\SI{30}{\electronvolt\squared}$ we observe negative slope values. For large $\SI{30}{\electronvolt\squared} \lesssim m_4^2 \lesssim \SI{1000}{\electronvolt\squared}$, the absolute magnitude of the slope is reduced and has the opposite sign. 
Due to the correlation between $m_4^2$ and $\mtwonue$, the exclusion curves vary significantly for different treatments of $\mtwonue$ as discussed in the following sections.  

\subsection{Neutrino mass fixed to \SI{0}{\electronvolt\squared}}\label{sec:result_fix}
\begin{figure}
\begin{center}
\includegraphics[width=\columnwidth]{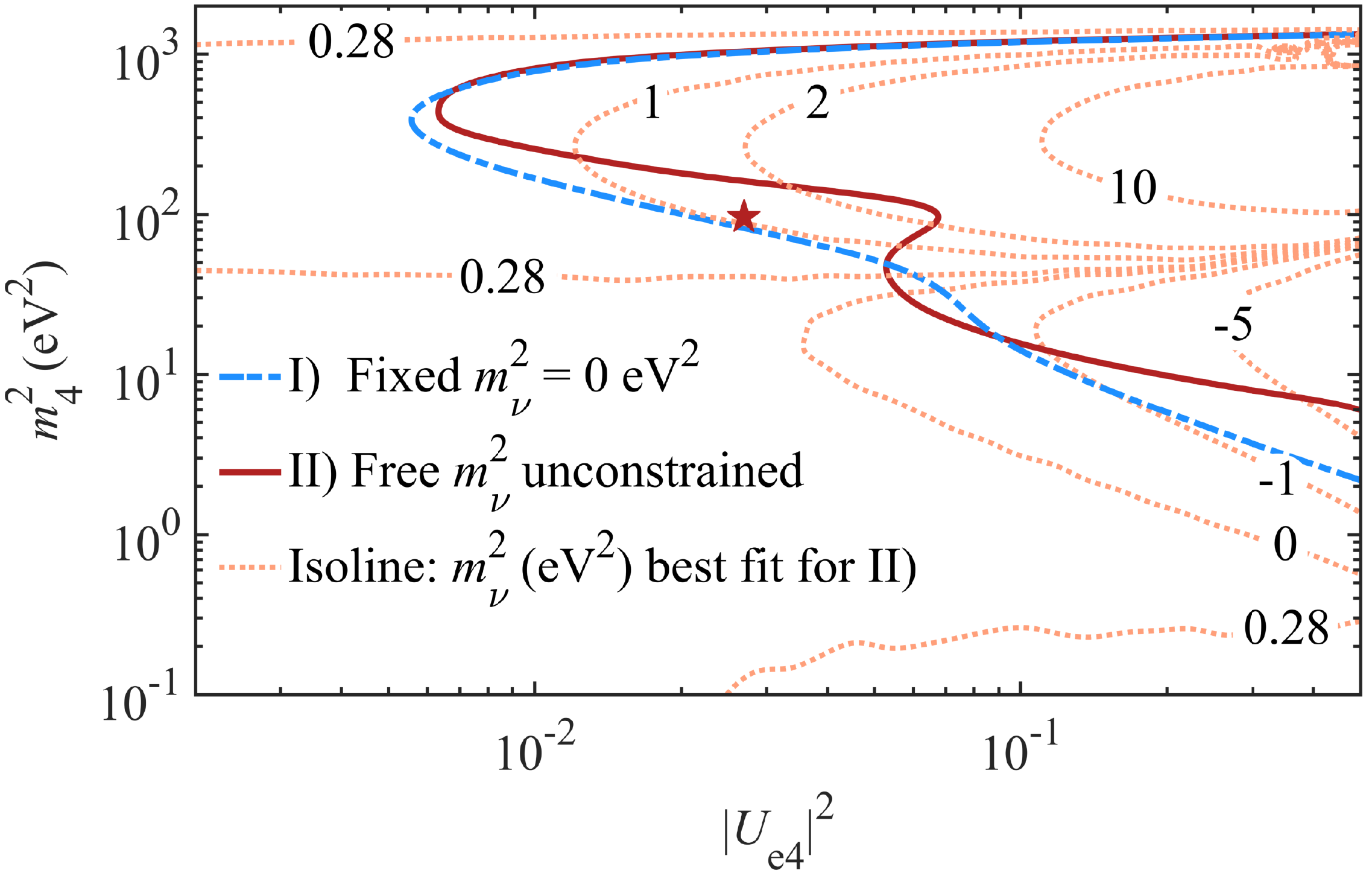}
\caption{\label{fig:Contour_mNuSqFree} Exclusion contours for data with \rom{1}) fixed $\mtwonue$ and \rom{2}) free $\mtwonue$. The $\mtwonue$ fit results within the grid search with free $\mtwonue$ are illustrated as isolines, i.e., $(m_4^2, |U_{e4}|^2)$ pairs with equal fit value of $\mtwonue$. The isolines are labelled with the corresponding value of $\mtwonue$ in \SI{}{\electronvolt\squared}. For small mixing $|U_{e4}|^2 < 6 \cdot 10^{-3}$, the neutrino-mass squared values lie within the $1\sigma$ confidence region of the neutrino-mass analysis~\cite{Aker:2021gma}, as expected. 
}
\end{center}
\end{figure}

In our main analysis, denoted case \rom{1}), we consider the hierarchical scenario $m_{1,2,3} \ll m_{4}$. This justifies setting $\mnue$ to zero, which is consistent with the lower limit derived from neutrino oscillations (\SI{0.009}{\electronvolt}~\cite{ParticleDataGroup:2020ssz}) within our sensitivity. We perform a two-dimensional grid search over $(m_4^2, |U_{e4}|^2)$, minimizing the $\chi^2$-function with respect to three free fit parameters ($E_0, A_\mathrm{S}, R_\mathrm{bg}^\mathrm{base}$) at each grid point. As $\mtwonue = \SI{0}{\electronvolt\squared}$ is fixed, we extend our grid to $|U_{e4}|^2= 1.0$. The exclusion curve at \SI{95}{\percent} C.L. is shown in Fig.~\ref{fig:Contour_mNuSqFree} (blue line). The global minimum of the $\chi^2$-function $\chi^2_\textrm{min}=27.5$ (23 dof, $p=0.24$) is found at $m_4^2=\SI{0.28}{\electronvolt\squared}$ and $|U_{e4}|^2=\SI{1.0}{}$.

The extreme active-to-sterile mixing shows that the observed decay spectrum (Eq.~\eqref{eq:diff-spectrum}) can be described best by only one branch with one associated free neutrino mass. As the active and sterile branches only differ in their neutrino mass and mixing nomenclature, the two branches are indistinguishable in the scenario at hand. However, since active-to-sterile mixing values of $|U_{e4}|^2 > \SI{0.5}{}$ are excluded by oscillation experiments~\cite{ParticleDataGroup:2020ssz}, we interpret our result as signature from the active branch with free $\mtwonue$. 
Indeed, the best-fit value for $m_4^2$ coincides with the best-fit value of $\mtwonue=\SI{0.28}{\electronvolt\squared}$ that was found in our neutrino-mass analysis with the same FPD pixel combination strategy~\cite{Aker:2021gma}. The significance of the best fit with respect to the null hypothesis is $\Delta\chi^2 = \chi^2_\text{null} - \chi^2_\mathrm{min} = 0.7$, i.e.,\ the result is not statistically significant ($\Delta\chi^2 < 5.99$).
Moreover, we perform a supplementary analysis, extending our modeled sterile-neutrino signal to the nonphysical parameter space: $m_4^2 \in [-\SI{40}{\electronvolt\squared},+\SI{40}{\electronvolt\squared}]$, $|U_{e4}|^2 \in [-0.5, 1]$. We find a global best fit for negative mixing and positive sterile-neutrino mass squared, which would formally correspond to a negative decay rate. 
While still not being significant at \SI{95}{\percent} C.L., this best fit improves the $\chi^2_\textrm{min}$ with respect to the minimum found in the physical region by one unit.
\subsection{Analysis with free neutrino mass}\label{sec:ResultFreemNuSq}
Furthermore, we study a more generic scenario, labeled case \rom{2}), in which we include $\mtwonue$ as an additional unconstrained nuisance parameter in our analysis. Due to the correlation between active and sterile neutrino mass (see Fig.~\ref{fig:NuMassCorrelation}), we expect a significantly different result than in analysis case \rom{1}. The associated exclusion contour (solid red) and the isolines of the $\mtwonue$ fit values (dotted light red) are displayed in Fig.~\ref{fig:Contour_mNuSqFree}. At each $(m^2_{4}, |U_{e4}|^2)$ pair in our grid, the fit has gained the freedom to converge to a value of $\mtwonue$ that improves the goodness-of-fit:
 \begin{equation}\label{eq:chi2imp}
 \begin{split}
 \chi^2_\textrm{min}(\mtwonue, m^2_{4}, |U_{e4}|^2) \\ \leq \chi^2_\textrm{min}(\mtwonue = \SI{0}{\electronvolt\squared}, m^2_{4}, |U_{e4}|^2).
\end{split}
\end{equation}
For small $m_4^2 \leq \SI{40}{\electronvolt\squared}$ and large $|U_{e4}|^2 \geq 0.04$, we find $\mtwonue \leq \SI{0}{\electronvolt\squared}$ of the same order of magnitude as $m_4^2$, reflecting the expected strong negative correlation between $\mtwonue$ and $m_4^2$.
For small mixing, $|U_{e4}|^2 < 6 \cdot 10^{-3}$, the $\mtwonue$ fit values lie within the $1\sigma$ confidence region of the neutrino-mass analysis~\cite{Aker:2021gma}. We report a best fit at $m_4^2=\SI{97.8}{\electronvolt\squared}$, $|U_{e4}|^2 = \SI{0.027}{}$ and $m_\upnu^2=\SI{1.1}{\electronvolt\squared}$ with $\chi^2 = 25.0$ (dof = 22, $p=0.30$).
The best fit improves with respect to the null hypothesis by $\Delta\chi^2=2.5$, thus not reaching the $\Delta\chi^2$ threshold at \SI{95}{\percent} C.L. for a significant result.
\subsection{\em Combined analysis} \label{sec:results_combi}
We combine the $3\nu+1$ sterile-neutrino constraints from this work with the result from the first four-week science run of KATRIN (KNM1). Since the first stand-alone analysis presented in~\cite{KATRIN:2020dpx}, the spectrum calculation has been slightly refined. It now incorporates the non-isotropic transmission of electrons,
a possible time-dependency of Penning trap induced background, an improved parametrization of the energy-loss function~\cite{KATRIN:2021rqj}, and reduced systematic uncertainties on the magnetic fields and the column density. The re-analysis of the first science run with these new inputs yields consistent exclusion contours to those of the original publication. 

The combined exclusion contours are obtained by minimizing the $\chi^2$-function
\begin{equation}
\label{eq:chi2combi}
\begin{split}
&\chi^2(\mtwonue, m_4^2, |U_{e4}|^2, \vec{\eta}_\text{KNM1}, \vec{\eta}_\text{KNM2}) = \\ &\chi^2(\mtwonue, m_4^2, |U_{e4}|^2, \vec{\eta}_\text{KNM1}) \, +\\ &\chi^2(\mtwonue, m_4^2, |U_{e4}|^2, \vec{\eta}_\text{KNM2})
\end{split}
\end{equation}
at each $(m_4^2, |U_{e4}|^2)$ pair. Due to different experimental conditions, several nuisance parameters are expected to vary between the data sets. As KNM2 was operated at a lower background level and a higher source activity compared to KNM1, we allow for a campaign-wise background and signal normalization. To account for an unknown difference in the absolute source potential, both data sets are described with an individual effective endpoint. The campaign-wise fit parameters ($E_0$, $A_\mathrm{S}$, $R_\mathrm{bg}^\mathrm{base}$) are summarized in Eq.~\eqref{eq:chi2combi} by $\vec{\eta}_\text{KNM1}$ and $\vec{\eta}_\text{KNM2}$ for the first and second measurement campaign. As both data sets are strongly statistics-dominated, possible correlations among systematic uncertainties are negligible.

\begin{figure}[htp]
\begin{center}
\includegraphics[width=\columnwidth]{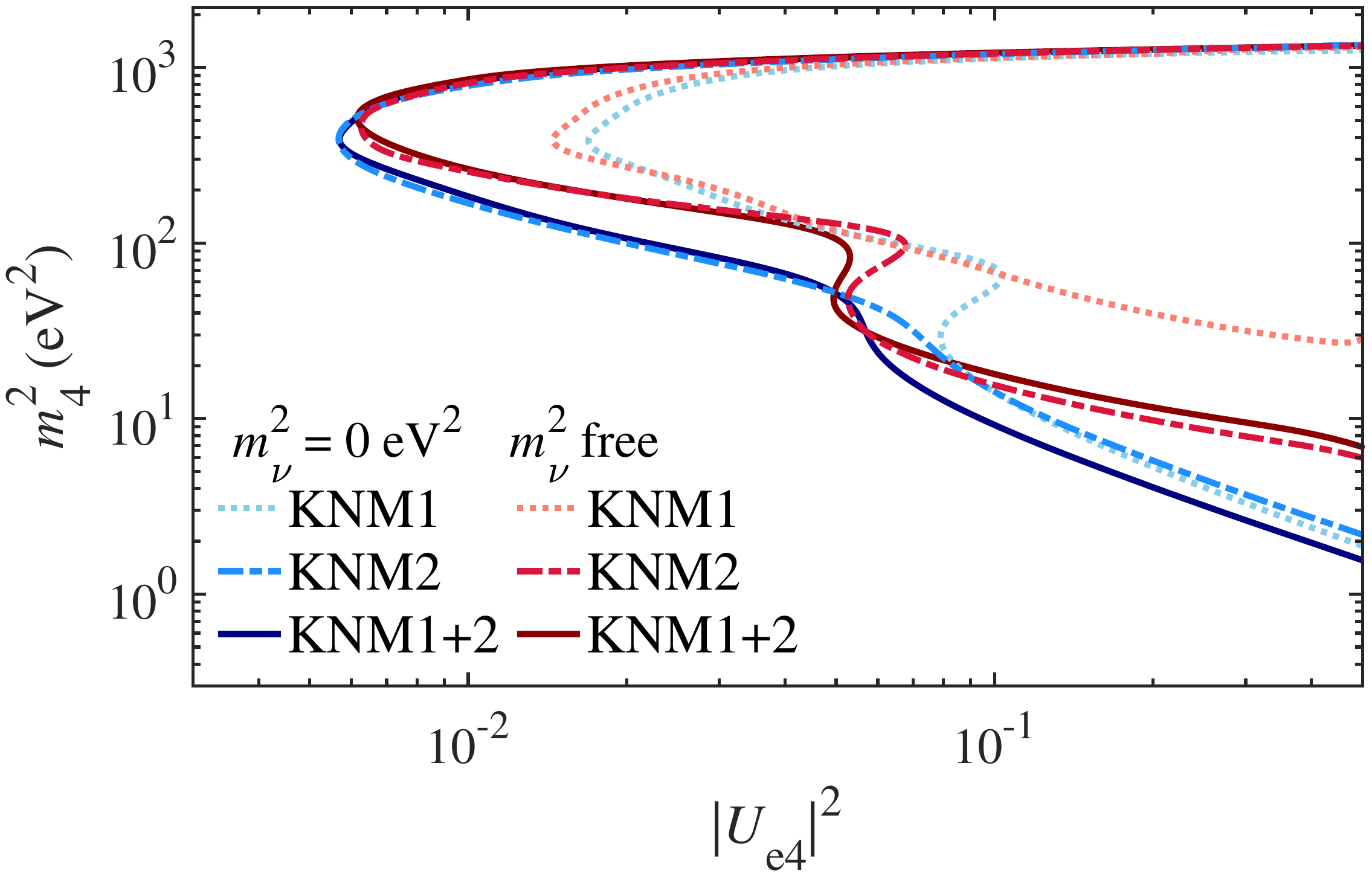}
\caption{\label{fig:Contour_Combi} Exclusion contours of the KNM1 and KNM2 standalone analyses and the combined analysis at \SI{95}{\percent} C.L. The contour lines shaded in blue consider $\mtwonue = \SI{0}{\electronvolt\squared}$, whereas the exclusions shaded in red include $\mtwonue$ as an unconstrained fit parameter. Due to statistical fluctuations, some contours exhibit bump-like features in the region $\SI{30}{\electronvolt\squared} \lesssim m_4^2 \lesssim \SI{100}{\electronvolt\squared}$. The location of the ($m_4^2$, $|U_{e4}|^2$) best-fit values, summarized in Tab.~\ref{tab:BestFitOverview}, generally cause slightly weaker constraints in their vicinity compared to the sensitivity. For the same reason, the combined exclusion contour of KNM1+2 gives a weaker constraint in certain regions compared to KNM2 standalone exclusion contour.}
\end{center}
\end{figure}

In case \rom{1}, KNM1 and KNM2 do not share any common nuisance parameter, because $\mtwonue$ is fixed. Therefore, the $\chi^2$-functions of the individual analyses in Eq.~\eqref{eq:chi2combi} can be minimized independently from each other. The combined and individual exclusion contours are shown in Fig.~\ref{fig:Contour_Combi} in blueish colors. The corresponding best-fit parameters are stated in Tab.~\ref{tab:BestFitOverview}. We find a best fit of the combined analysis at ($m_4^2=\SI{59.9}{eV^2}$, $|U_{e4}|^2=\num{0.011}$) with $\chi^2_\textrm{min} = 50.4$ (dof = 47, $p=0.34$), improving with respect to the null hypothesis by $\Delta\chi^2 = 0.7$. To evaluate the compatibility between the two statistically independent data sets, we perform the Parameter-Goodness-of-Fit (PGoF) test~\cite{Maltoni:2003cu}. This test quantifies the penalty of combining KNM1 and KNM2 in units of $\chi^2$ compared to the stand-alone analyses
\begin{equation}
\chi^2_\mathrm{penalty} = \chi^2_\mathrm{min, combi.} -  (\chi^2_\mathrm{min, KNM1} + \chi^2_\mathrm{min, KNM2}).
\end{equation}
As the data sets share two fit parameters ($m_4^2$ and $|U_{e4}|^2$), the $\chi^2$-penalty can be converted into a $p$-value using two degrees of freedom. We report on a probability of $\hat{p} = \SI{47}{\percent}$, demonstrating a good agreement. Due to statistical fluctuations in both data sets, the combined exclusion improves for $m_4^2 \leq \SI{50}{\electronvolt\squared}$ compared to the KNM2 stand-alone result while providing slightly weaker constraints for $m_4^2 > \SI{50}{\electronvolt\squared}$. The KNM1 exclusion bounds are improved by the combined analysis for the entire mass range. The observed exclusion contour agrees well with our sensitivity estimate, lying within the $1\sigma$ band of \SI{95}{\percent} C.L. sensitivity contours that are obtained from the simulation of 1500 randomized pseudo-experiments.

In case \rom{2}, KNM1 and KNM2 share $\mtwonue$ as a common nuisance parameter. Therefore a simultaneous grid search on both data sets minimizing the combined $\chi^2$-function in Eq.~\eqref{eq:chi2combi} has to be performed. The exclusion contours of the standalone and combined analyses are displayed in Fig.~\ref{fig:Contour_Combi} with reddish coloring. The relevant parameters of the best fits are given in Tab.~\ref{tab:BestFitOverview}.
The best fit of the combined exclusion at $\chi^2 = 49.9$ (46 dof, $p = 0.34$) improves with respect to the null hypothesis by $\Delta \chi^2 = 1.7$, rendering it not significant at \SI{95}{\percent} C.L. The PGoF of \SI{20}{\percent} indicates a good compatibility between the KNM1 and KNM2 sterile-neutrino analysis with free $\mtwonue$. We find the best-fit value $\mtwonue = \SI{0.57}{\electronvolt\squared}$, which agrees within $1\sigma$ with the standard neutrino-mass analysis~\cite{Aker:2021gma}.
\begin{table*}[!ht]
\small
    \centering
   \caption{Results of the KNM1 and KNM2 (Sec.~\ref{sec:result_fix}, \ref{sec:ResultFreemNuSq}) standalone and the combined analyses (Sec.~\ref{sec:results_combi}). The first three rows correspond to analysis case \rom{1} with $\mtwonue = \SI{0}{\electronvolt\squared}$, whereas the last three rows show the results of analysis case \rom{2} with unconstrained $\mtwonue$. The first five columns show the best-fit parameter values ($m_4^2$, $|U_{e4}|^2$, $\mtwonue$) and the associated goodness-of-fits ($\chi^2_\mathrm{bf}$, $p$). Furthermore, the two following columns state the significance of the best fit over the no-sterile hypothesis in terms of $\chi^2$ and confidence level. All observed sterile-neutrino signals are compatible with the no-sterile neutrino hypothesis, i.e.\ no significant spectral distortions at \SI{95}{\percent} C.L. are found. The last column gives the parameter goodness-of-fit $\hat{p}$ for the combined analyses.}
\label{tab:BestFitOverview}
    \begin{tabular}{ccccccccc|c}
    \toprule
         Analysis case     & Data set &$m_4^2$& $|U_{e4}|^2$  & $m_\nu^2$&$\chi^2_\textrm{min} / $dof  &$p$ & $\Delta\chi^2_\textrm{null}$ &  Significance &$\hat{p}$ \\
    \midrule 
    \multirow{2}{*}{\rom{1}} &   KNM1      & \SI{77.5}{\electronvolt\squared} & 0.031 & fixed & 21.4/22  & 0.50 & 1.43 &\SI{51.0}{\percent}& -\\
     &   KNM2     & \SI{0.28}{\electronvolt\squared}& 1.0& fixed & 27.5/23 & 0.24 & 0.74 & \SI{31.0}{\percent} & - \\
     &  KNM1$+$2 & \SI{59.9}{\electronvolt\squared}& 0.011 & fixed & 50.4/47  & 0.34 & 0.66 & \SI{28.1}{\percent} & 0.47\\
        \midrule
      \multirow{2}{*}{\rom{2}} &    KNM1 & \SI{21.8}{\electronvolt\squared} & 0.155 & \SI{-5.3}{\electronvolt\squared} & 19.9/21  & 0.53 & 1.30 & \SI{47.9}{\percent} & -\\
     &   KNM2 & \SI{98.3}{\electronvolt\squared}& 0.027& \SI{1.1}{\electronvolt\squared}& 25.0/22  & 0.30 & 2.49 &  \SI{71.2}{\percent} & - \\
     &   KNM1$+$2 & \SI{87.4}{\electronvolt\squared} & 0.019 & \SI{0.57}{\electronvolt\squared}& 49.5/46 & 0.34 & 1.69 & \SI{57.1}{\percent} &0.20\\
        \bottomrule
    \end{tabular}
\end{table*}
\subsection{Neutrino-mass sensitivity}\label{sec:result_numass}
As described in Sec.~\ref{sec:NuMassCorrelation}, we observe a sizable correlation between the effective electron antineutrino mass and the fourth mass eigenstate. This relation results in weaker constraints on the active-to-sterile neutrino mixing when $m_\nu^2$ is included as a free fit parameter in the sterile-neutrino search (see Sec.~\ref{sec:ResultFreemNuSq}). Turning the analysis concept upside down, the $3\nu+1$ model extension is expected to cause a reduction in $\mtwonue$ sensitivity~\cite{Riis_2011}. 

\begin{figure}[htp]
\begin{center}
\includegraphics[width=\columnwidth]{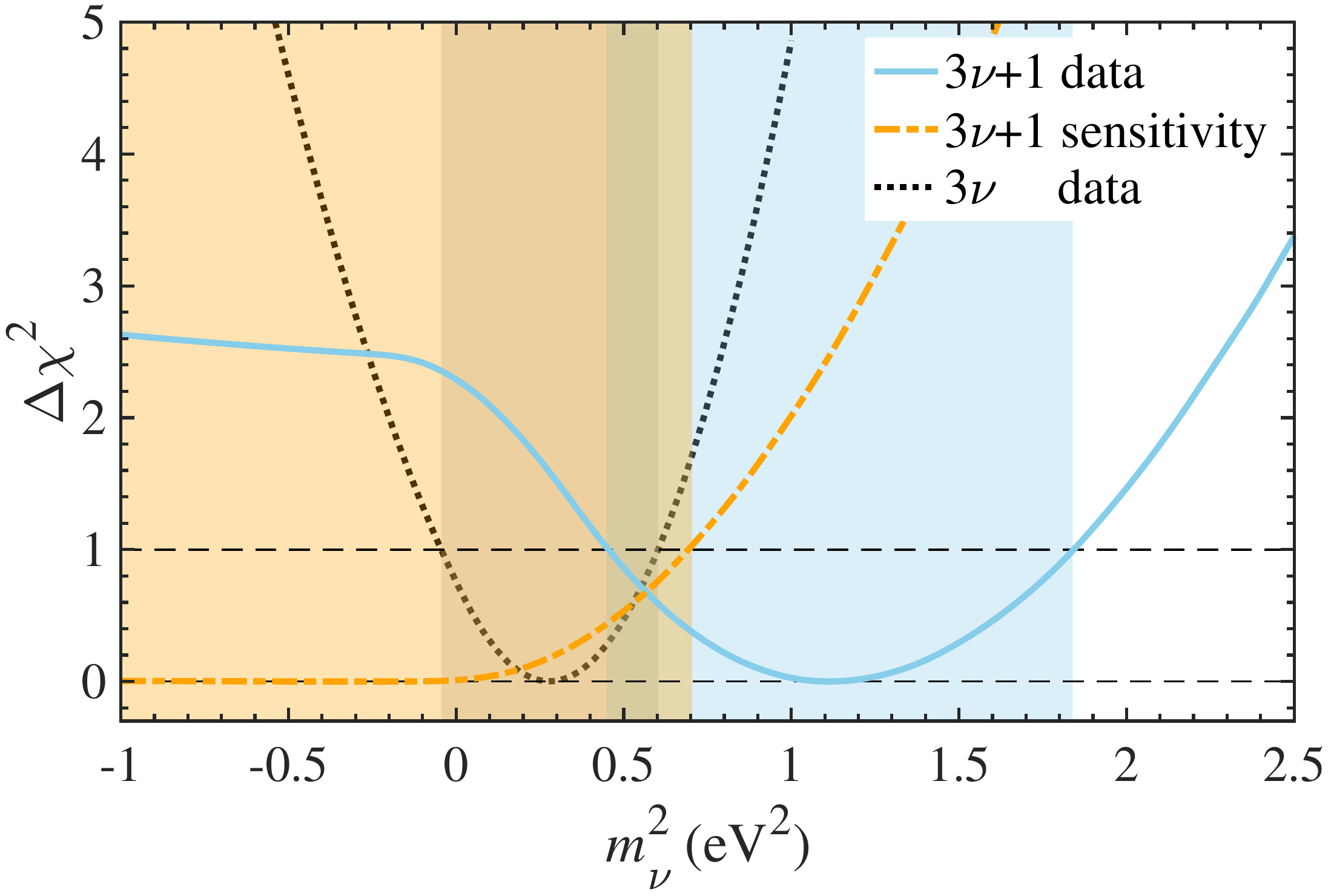}
\caption{\label{fig:NuMassSensitivity} Central value and $1\sigma$ uncertainty on $m_\nu^2$ within $3\nu+1$ framework for data (blue) and simulation (orange). The uncertainty obtained within the $3\nu$ framework is given in grey for comparison. }
\end{center}
\end{figure}

To assess the latter, we calculate the $\chi^2$-profile as a function of $\mtwonue$, displayed in Fig.~\ref{fig:NuMassSensitivity} for both data and twin analysis. For different fixed $\mtwonue \in [-1,2.5]$ \si{\electronvolt\squared}, a two-dimensional grid search over the $(|U_{e4}|^2, m_4^2)$ parameter space is performed, minimizing the $\chi^2$-function with respect to all other nuisance parameters. The value $\chi^2(\mtwonue)$ in the $\chi^2$-profile corresponds to the global minimum found in the grid search with the respective fixed $\mtwonue$. 
For $m_\nu^2 \leq \SI{0}{\electronvolt\squared}$, the global $\chi^2$ minima are located at $m_4^2<\SI{2}{\electronvolt\squared}$ and large mixing $|U_{e4}|^2\approx 0.5$. As $m_\nu^2$ and $m_4^2$ are strongly correlated in this region (see Fig.~\ref{fig:NuMassCorrelation}), the $\chi^2$-profiles are flat. Assuming the existence of a sterile neutrino, this corresponds to a complete loss of sensitivity. The latter can be restored by using external constraints on $m_4^2$ or $|U_{e4}|^2$.  Restricting $m_4^2 > \SI{20}{\electronvolt\squared}$ or $|U_{e4}|^2 < 0.04$, we find lower and upper $1\sigma$ sensitivities on $m_\nu^2$ of equal size. For $m_\nu^2 > \SI{0}{\electronvolt\squared}$, the best fits are located at $\SI{10}{\electronvolt\squared} < m_4^2 < \SI{200}{\electronvolt\squared}$ and moderate mixings $\mathcal{O}(10^{-2})$. In this part of the parameter space, the correlation between the two masses is less pronounced. As a result, the $1\sigma$ uncertainty on $m_\nu^2$ in the $3\nu+1$ extension is only increased by a factor of $2$ compared to the standard neutrino-mass analysis.

The neutrino-mass sensitivity in the $3\nu+1$ framework can be fully restored by limiting the active-to-sterile mixing to small values. For $|U_{e4}|^2<10^{-4}$, the sensitivity on $m_\nu^2$ converges to the nominal one in the $3\nu$ framework. Using the same constraint, we can also reproduce the central value and uncertainties of our standard neutrino-mass analysis~\cite{Aker:2021gma}.
\section{Comparison to other experiments}
\begin{figure*}[htp]
\begin{center}
\includegraphics[width=2\columnwidth]{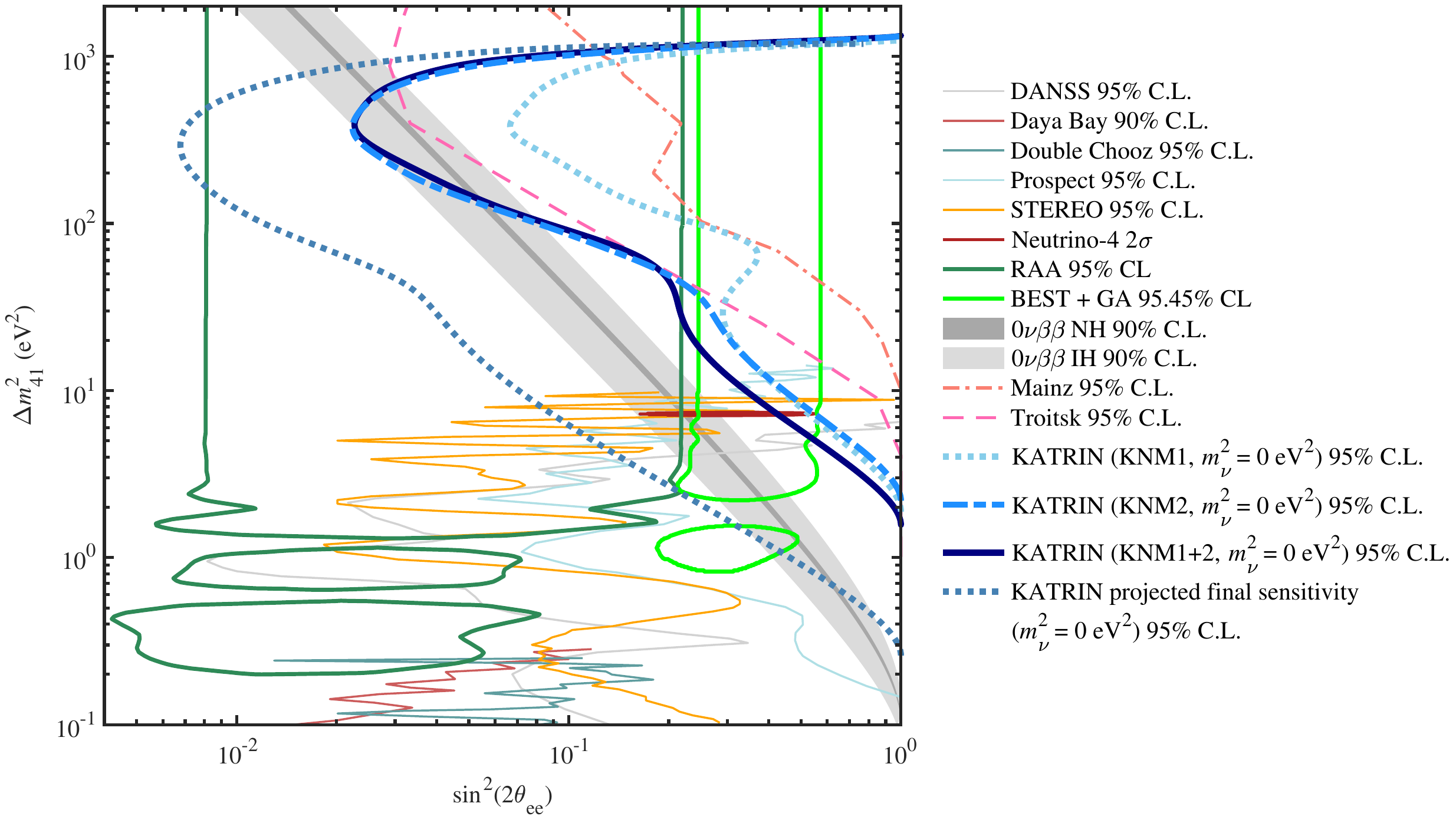}
\caption{\label{fig:osci} The \SI{95}{\percent} C.L. KATRIN exclusion contours from the first two measurement campaigns with $\mtwonue = \SI{0}{\electronvolt\squared}$, standalone and combined, are shown. The final sensitivity was computed following the first measurement campaign while assuming 1000 live days and a reduced background of \SI{130}{mcps}~\cite{KATRIN:2020dpx}. Our second measurement campaign yields more stringent constraints than both Mainz~\cite{Kraus:2012he} and Troitsk~\cite{Belesev:2012hx} experiments for $m_4^2\lesssim\SI{300}{\electronvolt\squared}$. We are able to exclude the large $\Delta m_{41}^2$ solutions of the RAA and BEST+GA anomalies~\cite{Mention:2011rk, barinov2021results} to a great extent. Our combined analysis is in tension with the positive results claimed by Neutrino-4~\cite{NEUTRINO-4:2018huq} for $\sin^2(2\theta_{ee}) \gtrsim 0.4$.
Moreover, KATRIN data improve the exclusion bounds set by short-baseline oscillation experiments for $\Delta m_{41}^2 \gtrsim \SI{10}{\electronvolt\squared}$~\cite{PROSPECT:2020sxr, Danilov:2019aef, MINOS:2020iqj, DoubleChooz:2020pnv, STEREO:2019ztb}. Constraints from $0\nu\upbeta\upbeta$ with $m_{\upbeta\upbeta}<\SI{0.16}{\electronvolt}$ are shown as gray bands~\cite{ParticleDataGroup:2020ssz,KamLAND-Zen:2016pfg,PhysRevLett.125.252502}.}
\end{center}
\end{figure*}
To put this work into context, we compare our case-\rom{1} exclusion contours with constraints from a selection of other experiments displayed in Fig.~\ref{fig:osci}, focusing on sterile neutrino searches in the electron disappearance channel. 
This result improves on the constraints from the completed Mainz and Troitsk experiments for  $m_4^2\lesssim\SI{300}{\electronvolt\squared}$.
As short-baseline neutrino oscillation experiments are sensitive to different observables than $\upbeta$-decay experiments, we perform the associated variable transformations to relate the results. While KATRIN is directly sensitive to $|U_{e4}|^2$, sterile neutrino oscillations are characterized by $\sin^2(2\theta_{ee}) = 4|U_{e4}|^2 (1-|U_{e4}|^2)$. Moreover, the mass splitting can be written as $\Delta m_{41}^2 \approx m_4^2 - m_\nu^2$, which is valid within $\SI{2E-04}{\electronvolt\squared}$~\cite{Giunti:2019fcj}.
For our analysis case \rom{1}, this approximation is equivalent to $\Delta m_{41}^2 \approx m_4^2$.
We are able to exclude the large $\Delta m_{41}^2$ solutions of the combined gallium experiments for $\SI{20}{\electronvolt\squared} \lesssim \Delta m_{41}^2 \lesssim \SI{1000}{\electronvolt\squared}$. Moreover, a considerable fraction of the reactor antineutrino anomaly for $\SI{50}{\electronvolt\squared} \lesssim \Delta m_{41}^2 \lesssim \SI{1000}{\electronvolt\squared}$  is challenged by our results. Our combined analysis of the first and second science run disfavors the Neutrino-4 hint of a signal for $\sin^2(2\theta_{ee}) \gtrsim 0.4$ at \SI{95}{\percent} C.L.

Furthermore, we compare our results to constraints from $0\nu\upbeta\upbeta$ experiments. If neutrinos are Majorana particles and $0\nu\upbeta\upbeta$ is triggered by light Majorana neutrino exchange, $m_4$ will contribute to the effective Majorana mass
\begin{equation}
\begin{split}
m_{\upbeta\upbeta} &= |\sum_{i=1}^{4} U^2_{ei} m_i| \\ &= |(1-|U_{e4}|^2) \sum_{i=1}^{3} U^2_{ei} m_i + |U_{e4}|^2 e^{i\gamma} m_4 | .
\end{split}
\end{equation}
with a possible third Majorana phase $\gamma\in[-\pi,\pi]$. Limiting the active neutrino contribution to $m_{\upbeta\upbeta}$ to the non-degenerate regime, i.e.\ between 0 and \SI{0.005}{\electronvolt} (0.01 and \SI{0.05}{\electronvolt}) for the normal (inverted) ordering, allows us to translate the current upper limit of $m_{\upbeta\upbeta}$ to constraints on sterile neutrinos. The conversion from the observable half-life to $m_{\upbeta\upbeta}$ depends on the nuclear matrix elements. As an illustration, we select the calculation with the nuclear matrix elements that result in the least stringent limit $m_{\upbeta\upbeta}<\SI{0.16}{\electronvolt}$~\cite{KamLAND-Zen:2016pfg,PhysRevLett.125.252502}. The width of the two gray exclusion bands in Fig.~\ref{fig:osci} reflects the uncertainties on the entries of the PMNS matrix and the unknown Majorana phases~\cite{ParticleDataGroup:2020ssz}. 

The projected final sensitivity quoted here after Ref.~\cite{KATRIN:2020dpx} demonstrates that KATRIN constraints will improve the global sensitivity for $\Delta m_{41}^2 \gtrsim \SI{5}{\electronvolt\squared}$ and will provide complementary results to short-baseline oscillation experiments for smaller masses.
\section{Conclusion}
We present the light sterile-neutrino search from the second KATRIN measurement campaign in 2019. Our data set comprises \SI{3.76E6}{} signal $\upbeta$-electrons inside the region of interest, reaching an energy-dependent signal-to-background ratio of up to $235$. The analysis is sensitive to the fourth neutrino mass eigenstate $m_4^2\lesssim\SI{1600}{\electronvolt\squared}$ and active-to-sterile mixing $|U_{e4}|^2 \gtrsim \SI{6e-3}{}$ in the $3\nu+1$ framework.
As no significant sterile-neutrino signal is observed, we report on improved exclusion limits with respect to our first measurement campaign. Our results improve on the constraints by previous tritium $\upbeta$-decay experiments. Moreover, we are able to exclude the large $\Delta m_{41}^2$ solutions of the reactor and gallium anomalies.
Combining the data sets from the first and second KATRIN measurement campaigns, our result disfavors the Neutrino-4 signal for $\sin^2(2\theta_{ee}) \gtrsim 0.4$.

The impact of systematic effects on our sterile-neutrino search was studied in detail. We conclude that our analysis is dominated by statistical uncertainties for all $m_4^2$ with a median relative contribution of $(\sigma^2_\mathrm{stat}/\sigma^2_\mathrm{total})_\mathrm{median} = \SI{86}{\percent}$ with respect to the total uncertainty budget. 

Furthermore, we investigated the correlation between active and sterile neutrino mass. We find a negative correlation for $m_4 \lesssim \SI{30}{\electronvolt\squared}$ with increasing absolute strength for increasing mixing. For larger sterile masses, the correlation is less pronounced and has a positive sign. Assuming the existence of a light sterile neutrino, this correlation translates into a reduction in neutrino-mass sensitivity by a factor of 2 compared to the neutrino-mass analysis in the $3\nu$ framework. By constraining the sterile neutrino mass or mixing, the nominal sensitivity can be restored. 

With hundreds of scheduled measurement days ahead, KATRIN will further improve its statistics by a factor of 50. In combination with a further reduction in background level and systematic uncertainties, this will allow us to cover an even larger fraction of the gallium and reactor antineutrino anomaly regions and the entire Neutrino-4 signal.
\begin{acknowledgments}
We acknowledge the support of Helmholtz Association (HGF), Ministry for Education and Research BMBF (05A20PMA, 05A20PX3, 05A20VK3), Helmholtz Alliance for Astroparticle Physics (HAP), the doctoral school KSETA at KIT, and Helmholtz Young Investigator Group (VH-NG-1055), Max Planck Research Group (MaxPlanck@TUM), and Deutsche Forschungsgemeinschaft DFG (Research Training Groups Grants No. GRK 1694 and GRK 2149, Graduate School Grant No. GSC 1085-KSETA, and SFB-1258) in Germany; Ministry of Education, Youth and Sport (CANAM-LM2015056, LTT19005) in the Czech Republic; Ministry of Science and Higher Education of the Russian Federation under contract 075-15-2020-778; and the Department of Energy through grants DE-FG02-97ER41020, DE-FG02-94ER40818, DE-SC0004036, DE-FG02-97ER41033, DE-FG02-97ER41041,  {DE-SC0011091 and DE-SC0019304 and the Federal Prime Agreement DE-AC02-05CH11231} in the United States. This project has received funding from the European Research Council (ERC) under the European Union Horizon 2020 research and innovation programme (grant agreement No. 852845). We thank the computing cluster support at the Institute for Astroparticle Physics at Karlsruhe Institute of Technology, Max Planck Computing and Data Facility (MPCDF), and National Energy Research Scientific Computing Center (NERSC) at Lawrence Berkeley National Laboratory.
\end{acknowledgments}
\bibliographystyle{apsrev4-2}
\bibliography{ksn2_compact}
\end{document}